\documentclass[%
 reprint,
superscriptaddress,
nofootinbib,
 amsmath,amssymb,
 aps,
]{revtex4-2}

\usepackage{graphicx}
\usepackage{dcolumn}
\usepackage{bm}
\usepackage{braket}
\usepackage{xcolor}
\usepackage{afterpage}
\usepackage{relsize}
\usepackage{multirow}
\usepackage{float}
\usepackage{physics}
\usepackage{makecell}
\usepackage{placeins}
\usepackage[colorlinks=true,urlcolor=blueprl,citecolor=blueprl,linkcolor=blueprl]{hyperref}
\newcolumntype{P}[1]{>{\centering\arraybackslash}p{#1}}

\definecolor{blueprl}{RGB}{46,48,146}

\def\ANU{Centre of Excellence for Quantum Computation and Communication Technology, The Department of Quantum Science and Technology, Research School of Physics and Engineering, The Australian National University, Canberra, Australian Capital Territory, Australia}

\def\ASTAR{Institute of Materials Research and Engineering, Agency for Science, Technology and Research (A*STAR), Singapore 138634}
\def\KIST{Center for Quantum Information, Korea Institute of Science and Technology (KIST), Seoul 02792, Republic of Korea}
\def\KISTNano{Division of Nano \& Information Technology, KIST School, Korea University of Science and Technology, Seoul 02792, Republic of Korea}

\begin{document}


\title{Surpassing the repeaterless bound with a photon-number encoded measurement-device-independent quantum key distribution protocol}

\author{\"{O}zlem Erk{\i}l{\i}\c{c}}
\email{ozlemerkilic1995@gmail.com}
\affiliation{\ANU}
\author{Lorc\'{a}n Conlon}
\affiliation{\ANU}
\author{Biveen Shajilal}
\affiliation{\ANU}
\author{Sebastian Kish}
\affiliation{\ANU}
\author{Spyros Tserkis}
\affiliation{\ANU}
\author{Yong-Su Kim}
\affiliation{\KIST}
\affiliation{\KISTNano}
\author{Ping Koy Lam}
\affiliation{\ANU}
\affiliation{\ASTAR}
\author{Syed M. Assad}
\email{cqtsma@gmail.com}
\affiliation{\ANU}

\date{\today}

\maketitle


\section*{Abstract}
Decoherence is detrimental to quantum key distribution~(QKD) over large distances. One of the proposed solutions is to use quantum repeaters, which divide the total distance between the users into smaller segments to minimise the effects of the losses in the channel. However, the secret key rates that repeater protocols can achieve are fundamentally bounded by the separation between each neighbouring node. Here we introduce a measurement-device-independent protocol which uses high-dimensional states prepared by two distant trusted parties and a coherent total photon number detection for the entanglement swapping measurement at the repeater station. We present an experimentally feasible protocol that can be implemented with current technology as the required states reduce down to the single-photon level over large distances. This protocol outperforms the existing measurement-device-independent and twin-field QKD protocols by surpassing the fundamental limit of the repeaterless bound for the pure-loss channel at a shorter distance and achieves a higher transmission distance in total when experimental imperfections are considered.

\section{\label{sec:level1}Introduction}
Quantum key distribution is a method used to securely establish a secret key between two distant trusted parties, namely Alice and Bob~\cite{ekert2014,gisin2002,pirandola2020advances}. Depending on the degrees of freedom of the underlying quantum system involved, QKD protocols are classified into two types, discrete-variable (DV) protocols where the key information is encoded on discrete degrees of freedom of photonic states such as polarisation~\cite{ch1984quantum,ekert1991quantum} and continuous-variable (CV) based protocols which encode the keys on continuous degrees of freedom such as amplitude and phase quadratures of the optical field~\cite{ralph1999,hillery2000}. In QKD, the main obstacle in establishing a secure key over large distances is the decoherence induced by photon losses.

Quantum repeaters are devices that can be used to improve the transmission distance of QKD protocols by dividing the total distance into smaller portions between the sender and receiver, making the losses in the channel more manageable~\cite{briegel1998quantum,dur1999quantum,duan2001long,Sangouard2011,munro2015inside}. Quantum repeaters~\cite{munro2015inside} use entanglement swapping~\cite{goebel2008multistage,kaltenbaek2009high,li2019experimental} to distribute entanglement, which is enhanced by entanglement distillation protocols~\cite{zhao2003experimental,vollbrecht2011entanglement,bratzik2013quantum}. One issue is that a majority of these repeater protocols require the use of quantum memories~\cite{simon2007quantum,Sangouard2011,dias2020}. However, quantum memories are limited by their operational wavelengths and memory efficiencies. Even though solid-state quantum memories \cite{bussieres2014quantum,stuart2021initialization} can operate at telecommunication wavelengths, their memory efficiency limits their efficacy. In contrast, cold-atom quantum memories currently hold the record for the efficiency, but operate outside of telecommunication wavelengths requiring frequency conversion to leverage communication infrastructure~\cite{cho2016highly,hsiao2018highly}. The frequency conversion results in low efficiencies limiting the performance of the current quantum repeaters~\cite{maring2014storage}.

The PLOB bound~\cite{pirandola2017fundamental} sets the fundamental limit for the maximum amount of private states that can be transferred in QKD for a given quantum channel without the use of a repeater (See Ref.~\cite{wilde2017converse} for the strong converse property of the bound and Ref.~\cite{pirandola2019end} for the bounds generalised to repeater-assisted communication). No point-to-point QKD protocol can surpass this bound unless there is a quantum repeater splitting the channel. Therefore, the PLOB bound can also be used as a benchmark to test the quality of quantum repeaters~\cite{pirandola2020advances}. It is known that the PLOB bound can be saturated with the squeezed-state protocol without the need for several copies of the states or a collective measurement for the pure-loss channel~\cite{pirandola2020advances}. When there is a repeater-chain, the end-to-end quantum capacity scales with the number of repeaters~\cite{pirandola2019end} and it is still an open question whether the corresponding repeater bounds can be saturated with a simple protocol without multiple copies of the quantum states.

Measurement-device-independent QKD (MDI-QKD) protocols are a type of repeater protocols in which the secret keys are established via the measurement of an untrusted third party~\cite{braunstein2012side,lo2012measurement,pirandola2013cvmdi,pirandola2015high}. These protocols are called \lq measurement-device-independent\rq\hspace{0.05cm} as Alice and Bob do not perform a measurement in their stations, but the measurement is performed by an untrusted party, called Charlie. Twin-field QKD (TF-QKD)~\cite{lucamarini2018} is a DV based MDI protocol which utilises weak identical coherent states sent by both Alice and Bob to Charlie, who performs entanglement swapping via a probabilistic photon detection measurement. TF-QKD protocol is the first repeater protocol without a quantum memory that is able to surpass the PLOB bound~\cite{lucamarini2018,chen511km,chen658km} as it scales proportionally to the single-repeater bound~\cite{pirandola2019end}. CV based MDI (CV-MDI) QKD protocols work in a similar fashion where Alice and Bob both send a distribution of either coherent or squeezed states to Charlie, where he performs a heterodyne measurement~\cite{pirandola2013cvmdi,pirandola2015high,wang2019cvmdi,ma2019cvmdi}.
In order to achieve a positive key rate in these CV-MDI protocols, the relay is positioned very close to Alice resulting in a very asymmetric set-up. As the relay is not placed right in the middle between Alice and Bob, the protocols scale like the repeaterless bound instead of the single-repeater bound. Hence, these protocols always sit below the PLOB bound.

In this work, we present a photon-number encoded MDI repeater protocol that surpasses the PLOB bound without the use of quantum memories through an entanglement swapping measurement. 
Unlike the TF-QKD protocol, the entanglement swapping is obtained by a coherent total photon number measurement performed by Charlie who measures the total number of photons coming from Alice and Bob without knowing the individual contributions. Even though the photon-number encoded states are vulnerable to losses, we show that in the short distance regime, the secret key rates are much higher than the ones of the single-photon encoded states. We also propose an experimentally feasible protocol using single-photons as these high dimensional states reduce down to the single-photon level over large distances. This protocol performs better than the existing MDI and TF-QKD protocols as it attains higher key rates for the same transmission distances.


\section{\label{sec:results}Results}
\subsection{\label{sec:protocol}The Measurement-Device-Independent Protocol}
\subsubsection{\label{sec:statessection}Alice and Bob's States for Generating a Key}
Let us assume that both Alice and Bob generate two-mode entangled states in their stations where they keep one arm of the entangled states to themselves and send the other to Charlie. Charlie then performs a joint entanglement swapping measurement on the states that Alice and Bob send.

QKD protocols can be expressed in either entanglement-based or prepare-and-measure schemes. Both of these models are mathematically equivalent~\cite{weedbrook2012,grosshans2003}, however the entanglement-based representation is more convenient for the security analysis of a QKD protocol. In the conventional entanglement-based CV-QKD protocols, Alice sends one arm of a two-mode squeezed vacuum state (TMSV) to Bob while performing a heterodyne measurement on the other arm of the TMSV state she kept. This procedure is equivalent to Alice sending a coherent state in the prepare-and-measure scheme~\cite{grosshans2003}. This entangled two-mode state in Fock basis is expressed as
\begin{equation}
\label{eq:eprstate}
\ket{\Psi}_{\mathrm{A_1 A_2}}=\frac{1}{\sqrt{N}}\sqrt{1-\gamma^2}\sum_{n=0}^{n_\text{max}}\gamma^n\ket{nn}_{\mathrm{A_{1}A_{2}}},
\end{equation}
where \text{$\gamma\in[0,1)$} is the squeezing parameter and $N$ is the normalisation coefficient given by $\sum_{n=0}^{n_\text{max}}(1-\gamma^2)\gamma^{2n}$. \text{$\ket{n}$} denotes the $n$-photon Fock state. Note that a TMSV state is retrieved with when \text{$n_\text{max}\rightarrow\infty$}~\cite{weedbrook2012}.
 
\begin{figure*}[htp!]
\center{\includegraphics[width=\textwidth]
        {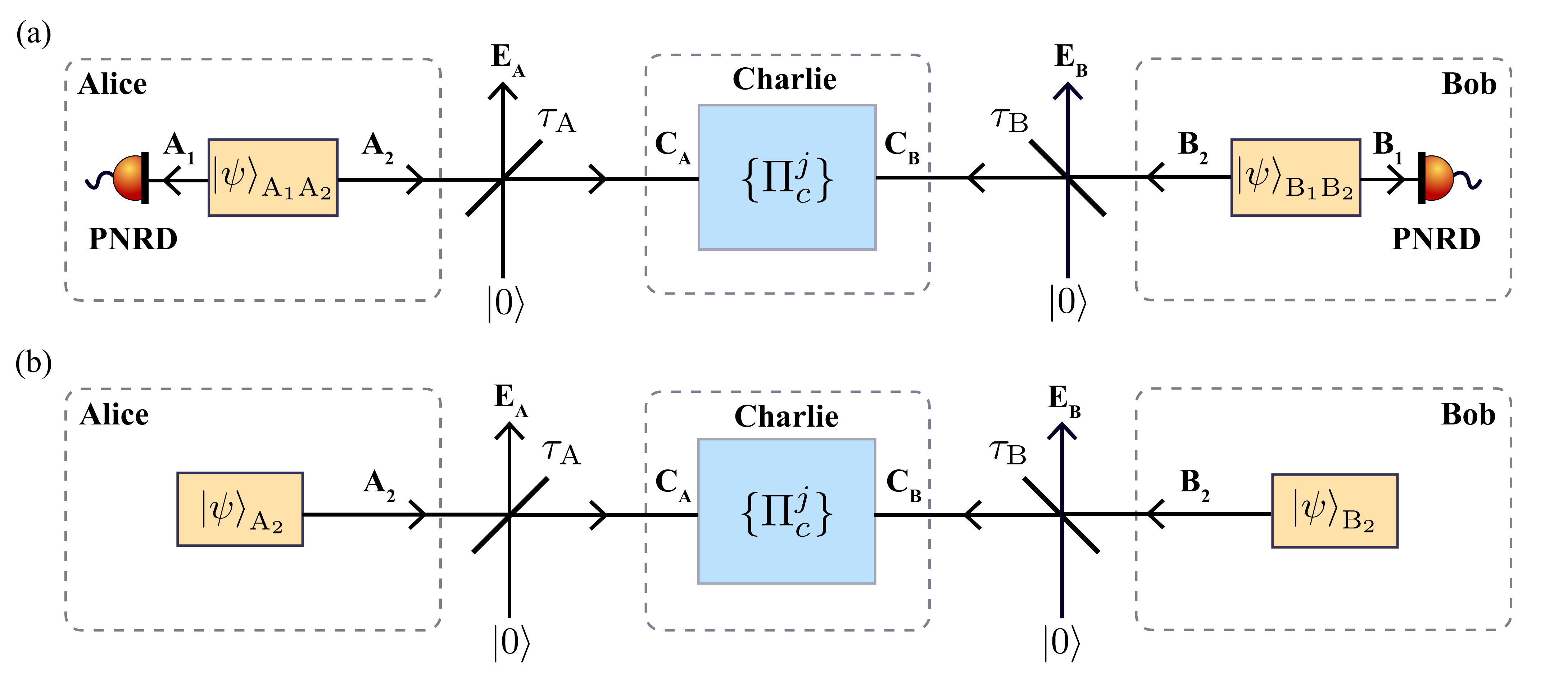}}
\caption{\label{fig:repeater_diagram} Equivalent representations of the protocol. (a) Entanglement-based scheme where both Alice and Bob send optimised states with \text{$n_\text{max}=7$} to Charlie while keeping one arm of their states to themselves denoted as modes A$_{\text{1}}$ and B$_{\text{1}}$ and measure the number of photons using photon-number resolving detectors (PNRDs). Charlie interferes modes A$_{\text{2}}$ and B$_{\text{2}}$ coming from Alice and Bob respectively and performs a coherent total photon number measurement. (b) Prepare-and-measure scheme where Alice and Bob send single-mode states to Charlie where they encode the key information on the single mode Fock state one at a time. Charlie then performs a total photon number measurement in his station on the modes that Alice and Bob send.}
\end{figure*}

In this paper, we use the entanglement-based version, shown in Fig.~\ref{fig:repeater_diagram}(a), for the security analysis of the prepare-and-measure method, shown in Fig.~\ref{fig:repeater_diagram}(b). We express Alice's and Bob's states as follows:
\begin{subequations}
\label{eq:statealice}
\begin{equation}
\label{eq:statealicesub}
\ket{\Psi}_{\mathrm{A_{1}A_{2}}}=\sum_{n=0}^{n_{\text{max}}}\sqrt{a_n}\ket{nn}_{\mathrm{A_{1}A_{2}}},
\end{equation}
\begin{equation}
\label{eq:bobsstatesub}
\ket{\Psi}_{\mathrm{B_{1}B_{2}}}=\sum_{n=0}^{n_{\text{max}}}\sqrt{b_n}\ket{nn}_{\mathrm{B_{1}B_{2}}},
\end{equation}
\end{subequations} 
where $\sum_{n=0}^{n_{\text{max}}}a_n=1$ and $\sum_{n=0}^{n_{\text{max}}}b_n=1$. \text{$a_{n}$} and \text{$b_{n}$} represent real coefficients of each Fock-number state \text{$\ket{nn}$}. These coefficients are the same for both Alice and Bob and optimised to achieve an optimal key rate explained in more detail in Sec.~\ref{sec:calkeyrate}. \text{$n_{\text{max}}$} is the maximum number of photons that Alice and Bob send individually, and each parties encode the key information on the Fock states \text{$\ket{n}$}. 

In the entanglement-based scheme, Alice and Bob keep one arm of the entangled states to measure the number of photons using a photon-number resolving detector (PNRD) to establish a key while sending the other arm to Charlie. Charlie performs a coherent total photon number measurement on the incoming modes from Alice and Bob, and announces the outcome of his measurement (described in detail in Sec.~\ref{sec:charliemeasurement}). Alice and Bob's measurement in their own stations is represented as
\begin{equation}
\Pi_{n}=\ketbra{n}{n},
\label{eq:pnrdpovm}
\end{equation}
where \text{$n$} denotes the number of photons being measured. 
In the prepare-and-measure scheme, this corresponds to preparing the Fock state $\ket{n}$ with probability $a_n$.
The states in the prepare-and-measure scheme can be engineered experimentally with several different methods such as conditional teleportation~\cite{asavanant2021wave}, coherent displacements and photon subtraction~\cite{fiuravsek2005conditional}, and repeated parametric-down conversion~\cite{clausen2001conditional}. Alternatively, these states can be created by extending the work presented in Ref.~\cite{bimbard2010quantum} to higher photon levels by using spontaneous parametric down-conversion on the signal channel and conditional measurements on the idler channel. 

\subsubsection{\label{sec:charliemeasurement}Charlie's Measurement}
The states are sent to Charlie via a channel with a total transmissivity of \text{$\tau\in[0,1]$}, which is split into smaller channels between Alice and Charlie and Charlie and Bob represented as \text{$\tau_{\mathrm{A}}$} and \text{$\tau_{\mathrm{B}}$} respectively. 
Single-repeater protocols can be benchmarked based on the PLOB bound which is given by \text{$-\mathrm{log}_{2}(1-\tau)$}~\cite{pirandola2017fundamental,pirandola2019end}. In order to surpass this bound, the protocol needs to scale like the single-repeater bound~\cite{pirandola2019end} which is expressed as \text{$-\mathrm{log}_{2}(1-\sqrt{\tau})$}. This requires Charlie to be positioned in the middle of Alice and Bob such that the key-rate scales with the square root of the transmission probability, $O(\sqrt{\tau})$. In this protocol, Charlie performs a collective photon number measurement on the incoming modes from Alice, \text{${\mathrm{A_2}}$}, and Bob, \text{$\mathrm{{B_2}}$}.
If Alice and Bob send a maximum of \text{$n$} photons each, denoted as \text{$n_\text{max}$}, Charlie can measure from $0$ to \text{$2n_\text{max}$} photons. Charlie's measurement can be realised by projecting the modes \text{${\mathrm{A_2}}$} and \text{$\mathrm{{B_2}}$} onto the following states
\begin{equation}
\ket{\phi_{c}^j}=\sum_{n=0}^{c}\frac{\omega^{nj}\ket{n}\!\ket{c-n}}{\sqrt{c+1}},
\label{eq:clickpovm}
\end{equation}
where \text{$c\in\{0,1,\cdots,2n_{\text{max}}\}$} represents the total number of photons Charlie receives from the two modes, and \text{$j\in\{0,1,\cdots,c \}$} denotes the different states in the $c$-photon subspace states while \text{$\omega$} is given by \text{$\omega=e^\frac{2\pi i}{c+1}$}. 

For example, when $c=2$, Charlie's three possible outcomes are
\begin{subequations}
\label{eq:alltheoutcomes}
\begin{equation}
\ket{\phi_{2}^0}=\frac{1}{\sqrt{3}}(\ket{02}+\ket{11}+\ket{20}),
\label{eq:phi0}
\end{equation}
\begin{equation}
\ket{\phi_{2}^1}=\frac{1}{\sqrt{3}}(\ket{02}+e^{\frac{2\pi i}{3}}\ket{11}+e^{-\frac{2\pi i}{3}}\ket{20}),
\label{eq:phi1}
\end{equation}
\begin{equation}
\ket{\phi_{2}^2}=\frac{1}{\sqrt{3}}(\ket{02}+e^{-\frac{2\pi i}{3}}\ket{11}+e^{\frac{2\pi i}{3}}\ket{20}).
\label{eq:phi2}
\end{equation}
\end{subequations}
These measurements are designed such that even though Charlie knows the total number of photons between Alice and Bob, he does not know the number of photons in each mode separately. 

The outcomes of Charlie's measurement form a valid positive operator value measurement (POVM) for a given outcome
\begin{equation}
\Pi_{c}^j=\ketbra{\phi_{c}^j}{\phi_{c}^j},
\label{eq:povm}
\end{equation}
with all the possible outcomes satisfying the identity resolution with  \text{$c\in\{0,1,\cdots,2n_{\text{max}}\}$} and \text{$j\in\{0,1,\cdots,c \}$}, i.e.,
\begin{equation}
\sum_{c=0}^{2n_\text{max}}\sum_{j=0}^c\Pi_{c}^j=\mathbb{I}.
\label{eq:povmidentity}
\end{equation}

The measurement performed by Charlie establishes correlations between Alice and Bob. In the lossless channel, when Charlie detects two photons with his POVM element \text{$\ket{\phi_{2}^0}$}, Alice and Bob's state becomes \text{$\ket{\psi}_{\mathrm{A_{1}B_{1}}|^{j=0}_{c=2}}=\sqrt{a_{0}a_{2}}\ket{02}+a_{1}\ket{11}+\sqrt{a_{2}a_{0}}\ket{20}$}. Therefore, Charlie swaps the entanglement between Alice and Bob via the measurement he performs similar to many MDI protocols~\cite{lo2012measurement,lucamarini2018}.

\subsubsection{\label{sec:checkstatessection}Alice and Bob's Check States for Security}
A possible security issue is that Charlie can potentially lie to Alice and Bob about his measurement outcome, as he can perform separable measurements on Alice and Bob's modes individually or announce a different photon number from the one he actually measured. When the latter occurs, Alice and Bob can tell that Charlie is not telling the truth as the probabilities of measuring different number of photons are not equal. However, when the former happens, Alice and Bob cannot distinguish whether Charlie is performing a total photon number measurement or a separable measurement on the two modes. Even though the separable measurement does not yield an entangled state between Alice and Bob, it still establishes classical correlations between the parties. The probability of Charlie measuring a given number of photons when he performs a separable measurement ends up being the same as his joint measurement described in Sec.~\ref{sec:charliemeasurement}.

We address this security issue by Alice and Bob randomly switching from their key states and sending some check states to Charlie to detect any abnormalities in the system. One of the possible check states they send consists of a superposition of the photon number states, and are analogous to the original DV diagonal states which are in the following form
\begin{equation}
\ket{+}=\frac{1}{\sqrt{n_{\text{max}}+1}}\sum_{n=0}^{n_{\text{max}}}\ket{n}.
\label{eq:diagonalstates}
\end{equation}

The untrusted party, Charlie, is required to announce the total number of photons he measured as well as the outcome index $(c,j)$. Table~\ref{tab:probtable} shows Charlie's probability of measuring $c=2$ photons as Alice and Bob send a mixture of key states and check states. Whenever both parties send $\ket{++}_{\mathrm{AB}}$, the probability of Charlie measuring $c=2$ photons is different for the non-separable and separable measurements. This is due to the nature of Charlie's POVM. For $c=2$, Charlie has three different outcomes in this set labelled as $\ket{\phi_{2}^0}$, $\ket{\phi_{2}^1}$, and $\ket{\phi_{2}^2}$. If Alice and Bob send $\ket{++}_{\mathrm{AB}}$, the probability of measuring $\ket{\phi_{2}^0}$ is $1/3$ whereas the other two outcomes return $0$. In the case of separable measurements, the probability of measuring a two-photon event is equal, allowing Alice and Bob to determine whether Charlie is being unfaithful or not.

\begin{table}[t!]
\renewcommand{\arraystretch}{1.2}
\caption{\label{tab:probtable}
Charlie's measurement probability for both non-separable and separable measurements for \text{$c=2$} when Alice and Bob send a combination of their check states, \text{$(+)$}, \text{$\ket{+}=\frac{1}{\sqrt{3}}(\ket{0}+\ket{1}+\ket{2})$} and key states, (\text{$K$}), \text{$\rho_{A_2}=\frac{1}{3}(\ketbra{0}+\ketbra{1}+\ketbra{2})$} in the prepare-and-measure representation with \text{$n_{\text{max}}=2$} photons.}
\begin{ruledtabular}
\begin{tabular}{cP{0.5cm}P{0.5cm}P{0.5cm}|P{0.5cm}P{0.5cm}P{0.5cm}}
 & \multicolumn{3}{c|}{Non-separable} &\multicolumn{3}{c}{Separable}\\
\hline
AB & \text{$\Pi_{2}^0$} & \text{$\Pi_{2}^1$} & \text{$\Pi_{2}^2$} & 02 & 11 & 20 \\
\colrule
$KK$ & 1/9 & 1/9 & 1/9 & 1/9 & 1/9 & 1/9\\
$K+$ & 1/9 & 1/9 & 1/9 & 1/9 & 1/9 & 1/9\\
$+K$ & 1/9 & 1/9 & 1/9 & 1/9 & 1/9 & 1/9\\
$++$ & 1/3 & 0 & 0 & 1/9 & 1/9 & 1/9\\
\end{tabular}
\end{ruledtabular}
\end{table}

The separable measurement is not the only possible measurement that Charlie can make. Ideally, Alice and Bob should not rely on Charlie's announcement of his measurement basis to determine if Charlie was being reliable or estimate how much information is leaked to another malicious party, called Eve. For security purposes, it is essential to utilise two or more non-orthogonal bases in QKD. For example, in BB84~\cite{ch1984quantum} and the six-state protocol~\cite{bruss1998optimal}, Alice sends states in two and three different orthogonal bases to Bob, respectively. By calculating the bit-error rates in these bases, Alice and Bob can estimate Eve's information. However, these protocols use only the probabilities of the matched measurement outcomes which overestimates Eve's information resulting in a lower key rate~\cite{liang2015tomographic}. Refs.~\cite{watanabe2008tomography,liang2015tomographic} showed that full tomography of the quantum state between Alice and Bob can enhance the secret key rate due to bounding Eve's information more accurately. Instead of using the statistics of the matched bases only, Alice and Bob can estimate their joint state from both the matched and unmatched bases. This joint state then can be used to calculate the Holevo bound on Eve's information. Holevo bound~\cite{holevo1998capacity} describes the maximum amount of classical information that can be extracted from a quantum channel. In QKD, Holevo bound can be used to upper bound the leaked information to Eve.

Our protocol requires a similar approach to the protocols discussed above~\cite{watanabe2008tomography,liang2015tomographic}, where Alice and Bob measure their joint state in mutually unbiased bases to perform a full tomography of their joint state in the entanglement-based scheme. Two bases $\{ \ket{e_i} \}_{i=0}^{m-1}$ and $\{ \ket{h_i} \}_{i=0}^{m-1}$ are called mutually unbiased when $|\braket{e_i}{h_j}|^2=1/m$ for any $i$ and $j$~\cite{schwinger1960unitary}, where \text{$m$} is the dimension of the Hilbert space. If the dimension of the Hilbert space, $m$, is a power of a prime number, there exists \text{$m+1$} mutually unbiased bases which form a complete set~\cite{wootters1989optimal}. In Methods~\ref{sec:appDVtomogprahy}, we show how to estimate Eve's information by reconstructing Alice and Bob's joint state through full tomography when Alice and Bob send single-photon states, i.e., \text{$n_\text{max}=1$}. In the entanglement-based scheme, Alice and Bob measure the modes they keep in their stations using the eigenvectors of the $X$, $Y$ and $Z$ bases which are expressed as
\begin{subequations}
\label{eq:allbases}
\begin{equation}
\ket{\pm x}=\frac{\ket{0}\pm\ket{1}}{\sqrt{2}},
\label{eq:xbasis}
\end{equation}
\begin{equation}
\ket{\pm y}=\frac{\ket{0}\pm i\ket{1}}{\sqrt{2}},
\label{eq:ybasis}
\end{equation}
\begin{equation}
\ket{+z}=\ket{0},\ \ket{- z}=\ket{1}.
\label{eq:zbasis}
\end{equation}
\end{subequations}
These bases form a complete set of mutually unbiased bases for \text{$m=2$}. In the equivalent prepare-and-measure scheme, Alice and Bob's measurement on the two mode entangled states, $\ket{\Psi}_{\mathrm{A_{1}A_{2}}}\!=\!\sqrt{a_0}\ket{00}+\sqrt{a_1}\ket{11}$ and $\ket{\Psi}_{\mathrm{B_{1}B_{2}}}\!=\!\sqrt{b_0}\ket{00}+\sqrt{b_1}\ket{11}$ in the $Z$ basis corresponds to them preparing the following states
\begin{subequations}
\label{eq:prepz}
\begin{equation}
\label{eq:st0}
\ket{\psi_{+z}}=\ket{0},
\end{equation}
\begin{equation}
\label{eq:st1}
\ket{\psi_{- z}}=\ket{1},
\end{equation}
\end{subequations}
with probability $a_0$ and $b_0$, and $a_1$ and $b_1$ respectively. Their measurement in the $X$ basis is equivalent to them preparing the following states with equal probability
\begin{equation}
\label{eq:pmx}
    \ket{\psi_{\pm x}}=\sqrt{\epsilon_0}\ket{0}\pm\sqrt{\epsilon_1}\ket{1},
\end{equation}
i.e., they prepare $\ket{\psi_{+x}}$ and $\ket{\psi_{-x}}$ with a probability of $0.5$, where $\epsilon_0$ and $\epsilon_1$ represent the coefficients $a_0$ and $b_0$, and $a_1$ and $b_1$, respectively. Similarly, their measurement in the $Y$ bases corresponds to them preparing the following states with equal probability
\begin{equation}
\label{eq:pmy}
    \ket{\psi_{\pm y}}=\sqrt{\epsilon_0}\ket{0}\mp i\sqrt{\epsilon_1}\ket{1}.
\end{equation}
We present the detailed results of this protocol in Sec.~\ref{sec:realimplementation}.

When Alice and Bob wish to encode the key onto the higher dimensional states, i.e., \text{$n_\text{max}>1$}, the number of check states they need to send increases. However, determining the existence of a complete set of mutually unbiased bases in an arbitrary dimensional Hilbert space is still an open problem in quantum information~\cite{horodecki2022five}. In this protocol, if Alice and Bob send states with \text{$n_\text{max}$} photons with a dimension of \text{$m=n_\text{max}+1$}, they need to send check states in \text{$m+1$} different bases to estimate Eve's Holevo bound provided that $m$ is a power of a prime number. These check states can be determined by following the method discussed in Ref.~\cite{wootters1989optimal}. We show the key rates of these higher dimensional states later in detail in  Sec.~\ref{sec:highdim} with \text{$n_\text{max}=7$} photons.

\subsubsection{\label{sec:calkeyrate}Calculation of the Secret Key Rate}
In the entanglement-based protocol, the global state before Charlie's measurement is a four-mode state. The dimension to simulate this protocol scales as \text{$m^4$}. Therefore, the coefficients of Alice and Bob's states in Eq.~\eqref{eq:statealice} are optimised by considering a classical protocol where Eve and Charlie perform a photon number measurement on their modes. We optimise the difference between the classical mutual information between Alice and Bob, and, Eve and Alice. We call this protocol the `classical protocol' and an explicit method for the implementation of this protocol is shown in the Methods~\ref{sec:appkeyclassical}. The reason for doing this is to avoid having to optimise a high dimensional four mode joint state with a total dimension of $m^4$. However, when computing the secret key rates, we do not assume any type of attacks for Eve and calculate Eve's Holevo bound instead and Charlie performs his collective photon number measurement. It is also important to note that the optimisation problem is not convex for the high-dimensional states and the solution provided for the coefficients $a_n$ and $b_n$ in this paper is one possible solution.

The states that Alice and Bob prepare are previously shown in Eq.~\eqref{eq:statealice}. They send these states through a pure-loss channel with a tranmissivity \text{$\tau_{\mathrm{A}}$} and \text{$\tau_{\mathrm{B}}$} for the channel between Alice and Charlie and Charlie and Bob, respectively. The pure-loss channel is modelled with a beamsplitter with a tranmissivity \text{$\tau$} where the beamsplitter mixes the input mode with the vacuum. The beamsplitter transformation can be defined as
\begin{equation}
B(\tau)=\mathrm{exp}[\mathrm{cos}^{-1}({\sqrt{\tau}})(\hat{a}^\dagger\hat{b}-\hat{a}\hat{b}^\dagger)],
\label{eq:beamsplitter}
\end{equation}
where \text{$\tau$} can be written as a function of the fibre distance, \text{$d$}, with a loss of 0.2dB per km with \text{$\tau=10^{-0.02d}$}. \text{$\hat{a}$} and \text{$\hat{b}$} are the annihilation operators, while \text{$\hat{a}^\dagger$} and \text{$\hat{b}^\dagger$} are the creation operators of the two modes respectively.

In this protocol, we assume that Eve has full access to the channel between Alice and Charlie and Charlie and Bob including Charlie's measurements. Eve mixes vacuum with the incoming modes causing Alice and Bob to lose photons. Thus, we can express the state between Alice and Charlie and Charlie and Bob after Eve's attack as
\begin{subequations}
\begin{equation}
\label{eq:alicesstate}
\rho_{\mathrm{A_{1}C_{A}}}\!=\!\mathrm{Tr_{3}}\!\big[\big\{\mathbb{I}_{m}{\otimes}B(\tau_\mathrm{{A}})\big\}\big\{\rho_{\mathrm{A_{1}A_{2}}}{\otimes}\!\ketbra{0}{0}\!\big\}\big\{\mathbb{I}_{m}{\otimes}B(\tau_{\mathrm{A}})\big\}^\dagger\big],
\end{equation}
\begin{equation}
\label{eq:bobsstate}
\rho_{\mathrm{C_{B}B_{1}}}\!=\!\mathrm{Tr_{1}}\!\big[\big\{B(\tau_{\mathrm{B}}){\otimes}\mathbb{I}_{m}\big\}\big\{\!\ketbra{0}{0}\!{\otimes}\rho_{\mathrm{B_{2}B_{1}}}\big\}
\\\big\{B(\tau_{\mathrm{B}}){\otimes}\mathbb{I}_{m}\big\}^\dagger\big],
\end{equation}
\end{subequations}
where \text{$\text{Tr}_i[\rho]$} stands for tracing out the $i$-th mode of the state \text{$\rho$}. 

After Charlie's measurement and tracing out his modes, the subnormalised state between Alice and Bob becomes
\begin{equation}
\label{eq:rhoAB}
\Tilde{\rho}_{\mathrm{AB}|^{j}_c}\!=\!\mathrm{Tr_{23}}\!\big[\big(\mathbb{I}_{m}{\otimes}\Pi_{c}^j{\otimes} \mathbb{I}_{m}\big)\!\big(\rho_{\mathrm{A_{1}C_{A}}}\!{\otimes}\rho_{\mathrm{C_{B}B_{1}}}\big)\!\big(\mathbb{I}_{m}{\otimes}\Pi_{c}^j{\otimes}\mathbb{I}_{m}\big)^\dagger\big].
\end{equation}

We can calculate Charlie's probability of obtaining outcomes \text{$(c,j)$} from the following expression
\begin{equation}
P_{^{j}_c}=\mathrm{Tr}\big[\Tilde{\rho}_{\mathrm{AB}|^{j}_c}\big].
\end{equation}

Normalising Alice and Bob's joint state by Charlie's probability of measuring \text{$c$} photons for his measurement \text{$j$} gives us the final conditional state between them as
\begin{equation}
\rho_{\mathrm{AB}|^{j}_c}=\frac{\Tilde{\rho}_{\mathrm{AB}|^{j}_c}}{P^{j}_{c}}.
\label{eq:rhoABnorm}
\end{equation}

However, for the key states that Alice and Bob send, the probability of Charlie measuring \text{$c$} photons, Alice and Bob's conditional mutual information and Eve's conditional information do not change for each \text{$j$} ranging from, \text{$0$} to \text{$c$}. As such, there is no need to calculate Alice and Bob's conditional joint state for each value of \text{$j$}. Therefore, we omit \text{$j$} from the following equations and set it to zero.

We then calculate Charlie's total probability of measuring \text{$c$} photons from
\begin{equation}
P_{c}=\sum_{j=0}^c\mathrm{Tr}\big[\Tilde{\rho}_{\mathrm{AB}|^{j}_c}\big]=(c+1)\mathrm{Tr}\big[\Tilde{\rho}_{\mathrm{AB}|^{j=0}_c}\big],
\end{equation}
since there are \text{$c+1$} POVM outcomes with a total photon number \text{$c$}.

In order to calculate Alice and Bob's mutual information, we first generate Alice and Bob's probability table as follows
\begin{equation}
P(n_{a},n_{b}|c)=\bra{n_{a},n_{b}}\!\rho_{\mathrm{AB}|_c}\!\ket{n_{a},n_{b}},
\label{eq:probABtable}
\end{equation}
where each term in Alice and Bob's mutual information is given by the conditional Shannon's entropy as expressed below
\begin{subequations}
\label{eq:alicebobiab}
\begin{equation}
H(A|c)=-\sum_{n_{a}\!=0}^{n_{\text{max}}}P(n_{a}|c)\log_{2}P(n_{a}|c),
\label{eq:subeqsalice}
\end{equation}
\begin{equation}
H(B|c)=-\sum_{n_{b}\!=0}^{n_{\text{max}}}P(n_{b}|c)\log_{2}P(n_{b}|c),
\label{eq:subeqsbob}
\end{equation}
\begin{equation}
H(AB|c)=-\sum_{n_{a}\!=0}^{n_{\text{max}}}\sum_{n_{b}\!=0}^{n_{\text{max}}}P(n_{a},n_{b}|c)\log_{2}P(n_{a},n_{b}|c).
\label{eq:subeeqsabtogether}
\end{equation}
\end{subequations}
Using the equations above, we evaluate Alice and Bob's mutual information conditioned on Charlie's measurement outcome from \text{$I_{AB|c}=H(A|c)+H(B|c)-H(AB|c)$}.

Eve's information is calculated from Alice and Bob's conditional state after Bob's measurement outcome on this joint state using
\begin{equation}
I_{E|c}=S(\rho_{\mathrm{AB}|c})-\sum_{b=0}^{n_{\text{max}}}P_{b}S(\rho_{\mathrm{A}|cb}),
\label{eq:eveinfo}
\end{equation}
where Bob's POVM is shown in Eq.~\eqref{eq:pnrdpovm} in Sec.~\ref{sec:protocol}. \text{$b$} represents the number of photons that Bob measures while \text{$P_{b}$} corresponds to Bob's probability of measuring \text{$b$} photons. The subnormalised state \text{$\Tilde{\rho}_{\mathrm{A}|cb}$} is obtained from
\begin{equation}
\Tilde{\rho}_{\text{A}|cb}=\mathrm{Tr}_{2}[(\mathbb{I}_{m}\otimes\Pi_{b})\rho_{\text{AB}|_c}(\mathbb{I}_{m}\otimes\Pi_{b})^\dagger],
\label{eq:evestate}
\end{equation}
where Bob's probability of measuring \text{$b$} photons is given by
\begin{equation}
P_{b}=\mathrm{Tr}[\Tilde{\rho}_{\mathrm{A}|cb}].
\end{equation}

Alice's subnormalised state conditioned on Bob's and Charlie's measurement outcomes, \text{$\Tilde{\rho}_{\mathrm{A}|cb}$} is then normalised by Bob's measurement probability by
\begin{equation}
\rho_{{\mathrm{A}|cb}}=\frac{\Tilde{\rho}_{\mathrm{A}|cb}}{P_{b}}.
\end{equation}

The asymptotic key rate of this protocol requires the combination of all the possible outcomes of Charlie's POVM since Alice and Bob are sending states with \text{$n$} photons each with a possibility of measuring \text{$0$} to \text{$2n$} photons by Charlie. However, we discard events where Eve's conditional information is greater than Alice and Bob's conditional mutual information. For example, when a zero photon occurs, Eve gets more information than Alice and Bob due to all the photons being lost to Eve. As such we exclude the case when \text{$c=0$}. Similarly, when Charlie measures \text{$c=2n_\text{max}$} photons, the key rate conditioned on this measurement outcome is zero even though Eve's conditional information is zero.
Therefore, the resulting asymptotic key rate can be expressed as
\begin{equation}
K=\sum_{c=0}^{2n_\text{max}}P_{c}\text{max}\big[0,I_{AB|c}-I_{E|c}\big].
\label{eq:keyrate}
\end{equation}

\begin{figure*}[htp!]
\includegraphics[width=\textwidth]{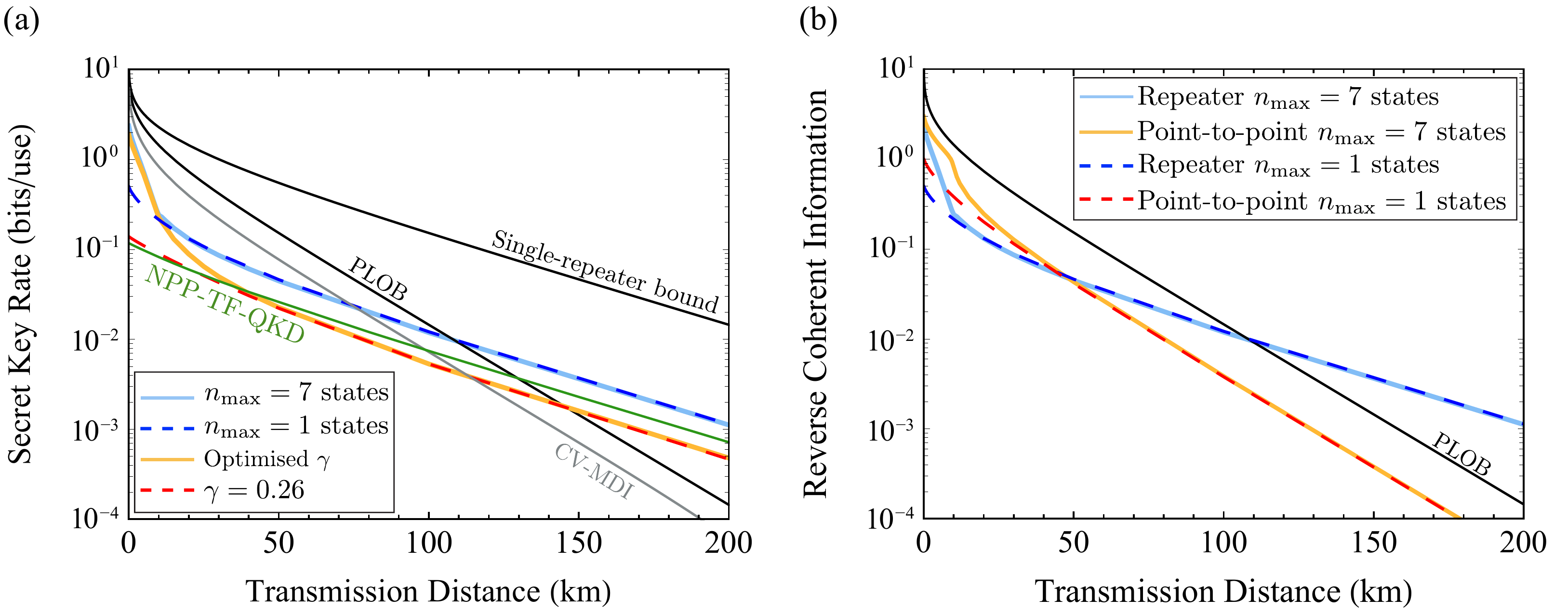}
\caption{\label{fig:repeater_plots}(a) Simulation results of our repeater protocol for a pure-loss channel with a loss of $0.2$dB/km. Solid orange and red dashed lines show our protocol using the states shown in Eq.~\eqref{eq:eprstate} with $n_\text{max}=7$ photons with optimised squeezing coefficients and a squeezing coefficient of \text{$\gamma=0.26$}, respectively. Blue dashed line shows the results of the optimised states with $n_\text{max}=1$ photon given in Eq.~\eqref{eq:statealice} and using Charlie's POVM with an outcome of $1$ and $2$ photons. The solid blue line shows our protocol using the optimised states with $n_\text{max}=7$ photons given in Eq.~\eqref{eq:statealice}. Black solid lines show the single-repeater and repeaterless bounds. Solid grey line represents the CV-MDI protocol~\cite{pirandola2013cvmdi,pirandola2015high} with a variance of 1000 and relay positioned at 0.01m away from Alice while the solid green line shows the TF-QKD protocol with no phase post-selection (NPP-TF-QKD) using optimised coherent states and infinite decoy states. (b) The comparison of the reverse coherent information of the optimised states with $n_\text{max}=7$ and $n_\text{max}=1$ photons in the form of Eq.~\eqref{eq:statealice} in point-to-point communications between Alice and Bob and with a single-repeater. The faint blue and orange lines represent the RCI of the single-repeater and point-to-point communications of the optimised states with $n_\text{max}=7$ photons correspondingly while the blue and red dashed lines show the RCI of the single-repeater and point-to-point communications of the optimised states with $n_\text{max}=1$ photon. The black solid line shows the reverse coherent information of an infinitely squeezed TMSV state denoted as PLOB.}
\end{figure*}

\subsection{\label{sec:highdim}The Results of the High-dimensional States}
Our simulation results are shown in Fig.~\ref{fig:repeater_plots}(a) for the pure-loss channel with $0.2$dB loss per km. We compare our results with the existing MDI protocols such as the CV-MDI protocol from Pirandola et al.~\cite{pirandola2013cvmdi,pirandola2015high} and one of the best performing TF-QKD protocols known as TF-QKD without phase post-selection (NPP-TF-QKD) from Cui et al.~\cite{cui2019twin} and Lu et al.~\cite{lu2019improving}. 

We first show the case where Alice and Bob send the states shown in Eq.~\eqref{eq:eprstate} with a squeezing coefficient of \text{$\gamma=0.26$} for each distance with \text{$n_{\text{max}}=7$} photons. The squeezing level of \text{$\gamma=0.26$} was determined based on the shortest distance that the protocol exceeds the PLOB bound (refer to Sec.~\ref{sec:appcoeffs} Table~\ref{tab:eprtable} for the details). With these states, the PLOB bound and the CV-MDI protocol are surpassed at \text{$144$} km and \text{$114$} km, respectively, while the protocol is performing worse than the TF-QKD protocol. We also demonstrate the key rates of the same states where the values of \text{$\gamma$} are optimised to give the maximum secret key rate at the corresponding distance. For distances greater than \text{$50$} km, there is not much difference compared to the states with \text{$\gamma=0.26$} and the PLOB bound is still surpassed at the same distance as the case of \text{$\gamma=0.26$}. However, the key rates are now higher at short distances below \text{$50$} km. This indicates that in the short distance regime, the contribution of the higher order photons to the key rate is significant while at larger distances, the main contribution comes from the the first few photons of the state as the majority of the photons are lost to the environment at such distances. This can also be seen from the optimal squeezing level given in Table~\ref{tab:eprtable}, which is higher for short distances and lower for larger distances.

\begin{figure*}[htp!]
\hspace*{-0.06cm} 
\includegraphics[width=\textwidth]{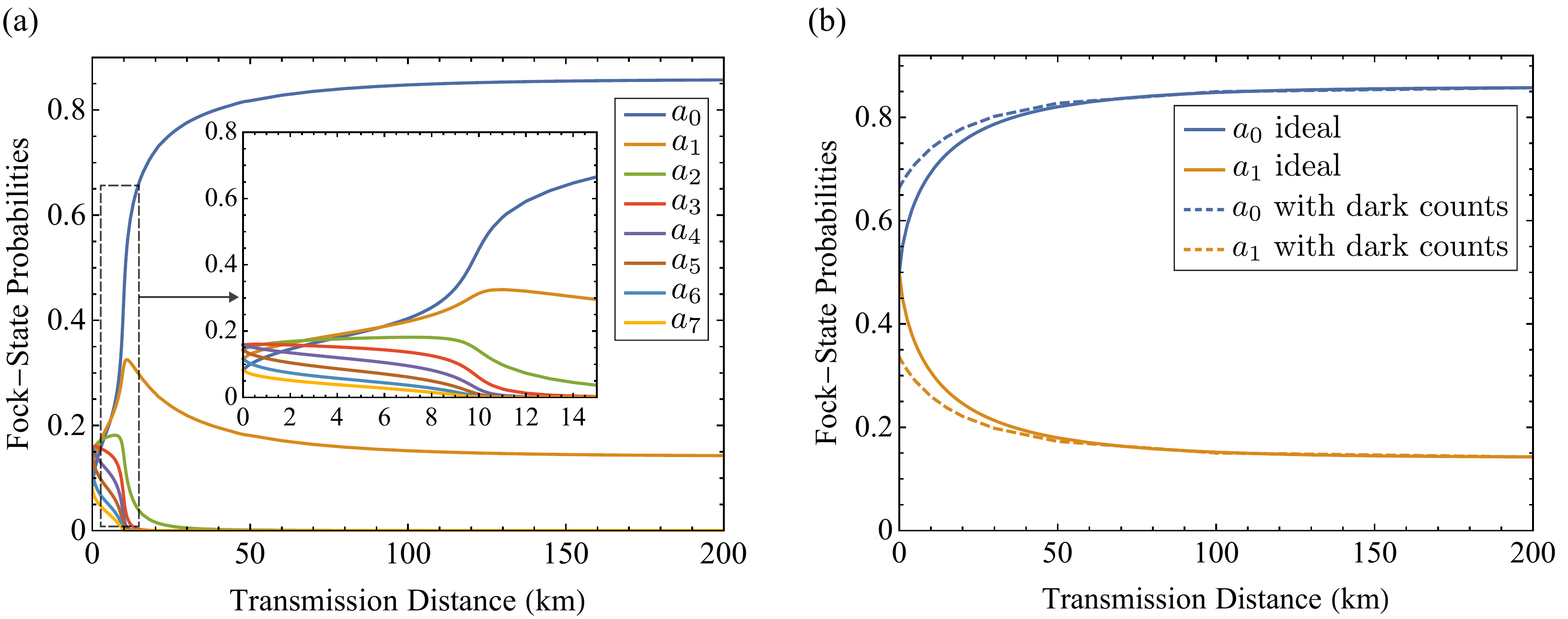}
\caption{\label{fig:opt_coefficients_figure}(a) The optimised coefficients of the states shown in Eq.~\eqref{eq:statealice} with $n_\text{max}=7$ photons, where the coefficients are explicitly shown in Table~\ref{tab:cvtable}. (b) The optimised coefficients of the states shown in Eq.~\eqref{eq:statealice} with $n_\text{max}=1$ photon (refer to Table~\ref{tab:dvtable} for the optimised coefficients for each distance).}
\end{figure*}

\begin{figure}[!b]
\hspace*{-0.5cm} 
\includegraphics[scale=0.56]{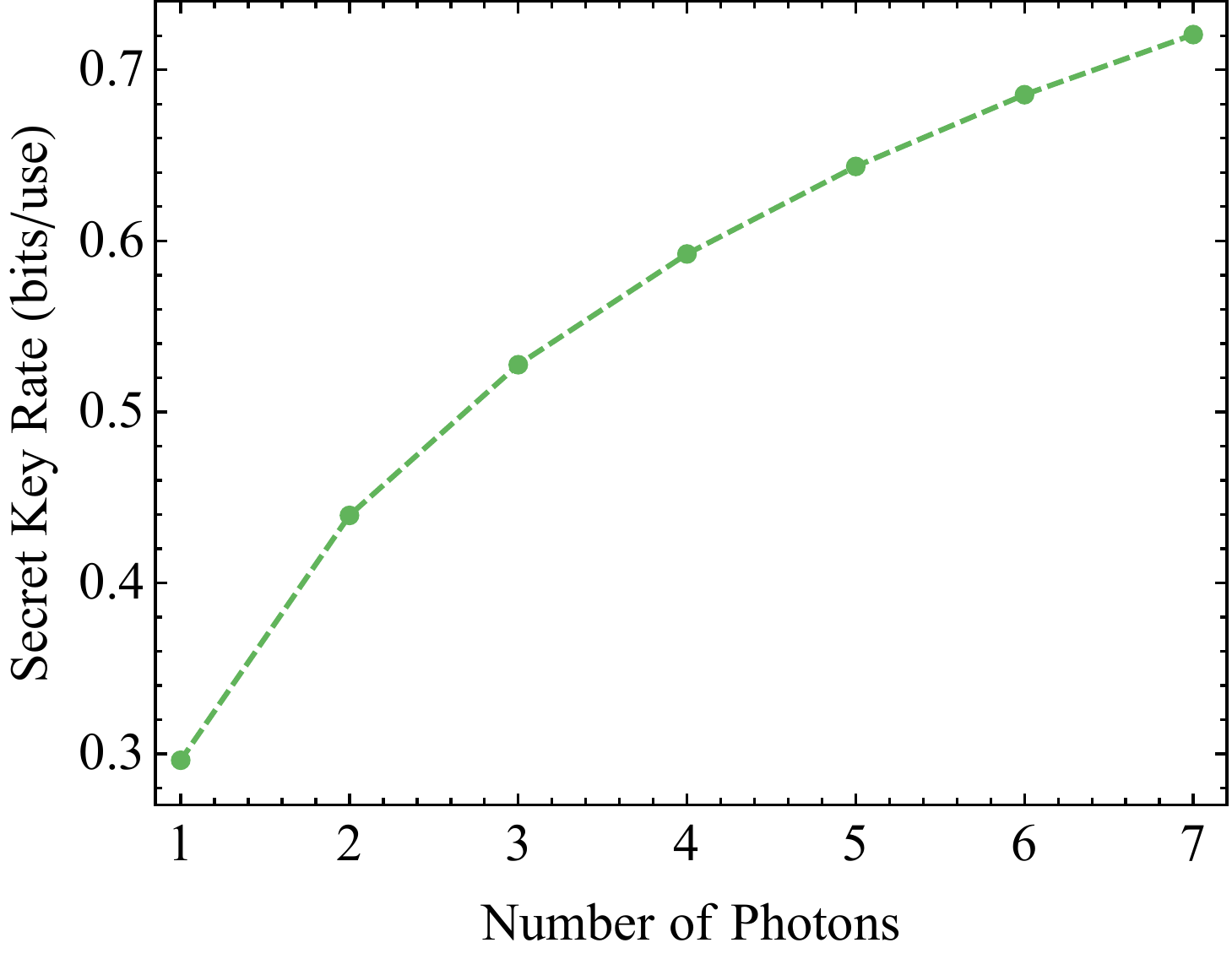}
\caption{\label{fig:different_photons} The simulation results of the secret key rate when the number of encoded photons varies from $1$ to $7$ photons at $5$ km. The states are in the form of Eq.~\eqref{eq:statealice} where the coefficients of each state are optimised.}
\end{figure}

When Alice and Bob send the optimised states shown in Eq.~(\ref{eq:statealice}), these states outperform the results of the states with optimised \text{$\gamma$} by surpassing the repeaterless bound and the CV-MDI protocol at \text{$108$} km and \text{$75$} km respectively. These states also do considerably better than the TF-QKD protocol as the TF-QKD protocol exceeds the PLOB bound at only \text{$130$} km and its key rates are lower than our protocol at each distance. It is important to note that this result can also be achieved by using the optimised states with \text{$n_\mathrm{max}=1$} photon in the form of \text{$\sqrt{a_{0}}\ket{00}+\sqrt{a_{1}}\ket{11}$} as shown in Fig.~\ref{fig:repeater_plots}(a) since both states reach the PLOB bound at the same distance and the key-rates converge beyond \text{$10$} km. In Fig.~\ref{fig:repeater_plots}(a), both high-dimensional and single-photons states have the same gradient, scaling like the single-repeater bound with $O(\sqrt{\tau})$. 
The probability of receiving $n$-photons in this case is given by $(\sqrt{\tau})^n$. Therefore, the main scaling of the key rates comes from the single-photon level while the remaining photons help the key rate incrementally. As the loss gets higher, the probability of receiving higher photons drops. Therefore, beyond $10$ km, we are only interested in \text{$1$} or \text{$2$} photons. This is further emphasised in Fig.~\ref{fig:opt_coefficients_figure}(a) where we show the probability of sending each Fock-number state of the optimised states given in Eq.~\eqref{eq:statealice} for each transmission distance. At short distances, the high-dimensional states have contribution from each photon number. It is important to note that at $0$ km, the probability of sending each Fock-number state is not equal due to key rate being equal to zero when Charlie receives $0$ or $14$ photons in total. Therefore, the coefficients of the Fock states $\ket{0}$ and $\ket{7}$ are minimised accordingly. As the distance increases, the high-dimensional states reduce down to the single-photon level as the coefficients of the Fock states above one photon approach zero. The probabilities of sending zero and one photon, denoted as \text{$a_0$} and \text{$a_1$}, of these high-dimensional states shown in Fig.~\ref{fig:opt_coefficients_figure}(a) converge to the coefficients of the optimised states with \text{$n_\mathrm{max}=1$} photon shown in Fig.~\ref{fig:opt_coefficients_figure}(b) beyond approximately $50$ km. However, the main advantage of using the optimised states with \text{$n_\mathrm{max}=7$} is the ability of obtaining higher key rates at shorter distances. This is shown in Fig.~\ref{fig:different_photons}, as the secret key rate increases when the number of encoded photons changes from $1$ to $7$ photons.

As the key rates of the optimised states with \text{$n_\mathrm{max}=7$} converge with the results of the states with optimised \text{$\gamma$} below $10$ km and with the optimised states with \text{$n_\mathrm{max}=1$} photon, one can use the combination of the states with optimised \text{$\gamma$} and optimised states with \text{$n_\mathrm{max}=1$} photon beyond this distance to achieve the same results of the states given in Eq.~(\ref{eq:statealice}).

As mentioned previously, the maximum key rate achievable by QKD for the point-to-point and single-repeater communication is bounded by the PLOB and single-repeater bounds respectively~\cite{pirandola2017fundamental,pirandola2019end}. These bounds are determined by the maximum amount of entanglement that a channel can sustain, also known as the entanglement flux, which coincides with reverse coherent information (RCI) of a maximally entangled TMSV state for the pure-loss channel~\cite{pirandola2017fundamental,pirandola2009direct,garcia2009reverse}. RCI is used to lower bound the distillable entanglement of a given channel~\cite{garcia2009reverse} and is a measure of the transmission of quantum information. While the key rates above demonstrate that our protocol surpasses the PLOB bound and acts as a repeater, the secret key rate is a measure of the transmission of classical information. The key rates are also bounded by the amount of entanglement that Alice and Bob can distill. Therefore, we also compute the RCI of our quantum states to verify the distillable entanglement between Alice and Bob after Charlie's measurement using
\begin{equation}
\text{RCI}=\sum_{c=0}^{2n_\text{max}}P_{c}\text{max}\bigl[0,S(\rho_{\text{A}|c})-S(\rho_{\text{AB}|c})\bigr],
\label{eq:rci}
\end{equation}
where \text{$S(\rho_{\text{AB}|c})$} and \text{$S(\rho_{\text{A}|c})$} are the von Neumann entropies of the joint state between Alice and Bob \text{$\rho_{\mathrm{AB}|c}$} and Alice's state \text{$\text{Tr}_2[\rho_{\text{AB}|c}]$} respectively. 
\begin{figure*}[htp!]
\hspace*{-0.2cm} 
\includegraphics[width=\textwidth]{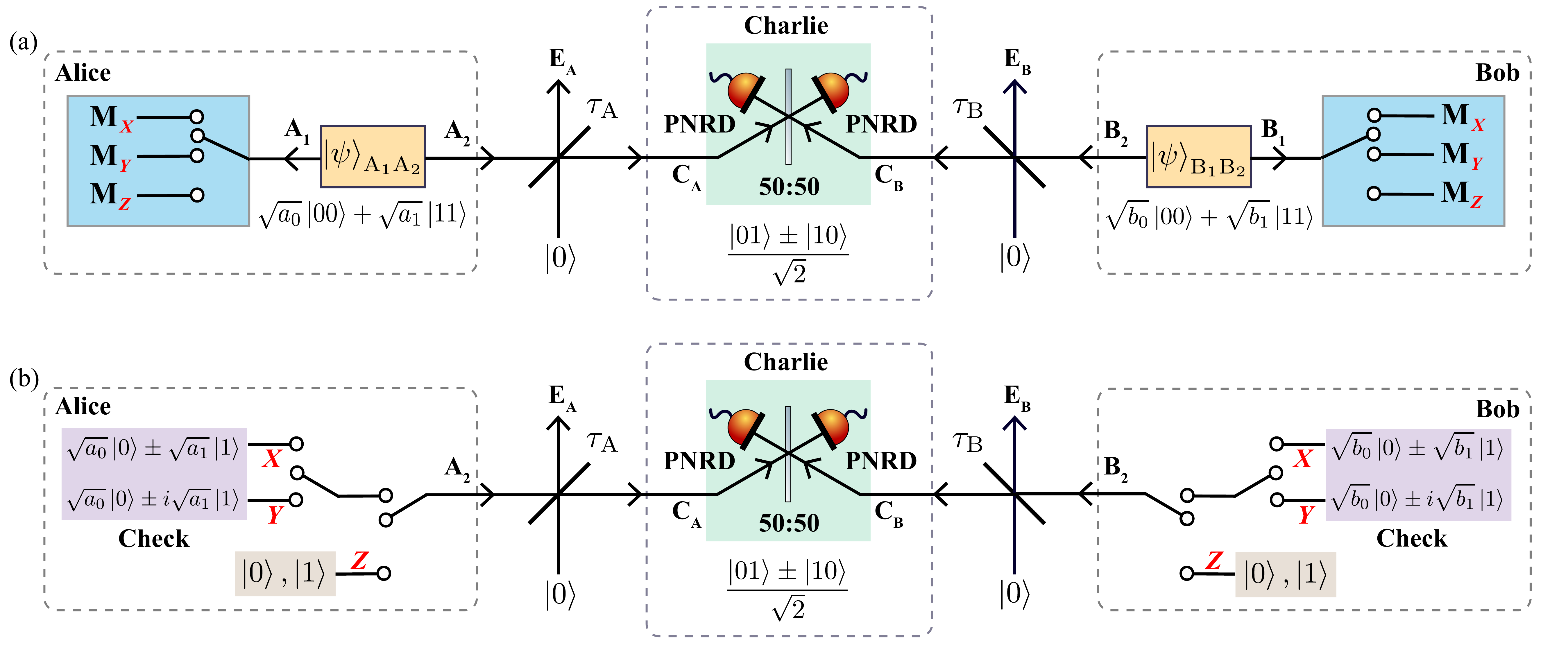}
\caption{\label{fig:single_photon_experiment}Equivalent representations of the protocol with single photons. (a) Entanglement-based scheme where both Alice and Bob send the optimised states in the form of \text{${\sqrt{a_0}}\ket{00}+{\sqrt{a_1}}\ket{11}$} and \text{${\sqrt{b_0}}\ket{00}+{\sqrt{b_1}}\ket{11}$} to Charlie while keeping one arm of their states to themselves denoted as modes A$_{\text{1}}$ and B$_{\text{1}}$. They perform a measurement in the $X$, $Y$ and $Z$ bases to compute the statistics of their matched and unmatched results. This measurement also projects the arm they sent to Charlie onto a single-mode state in the corresponding basis as shown in panel (b). Charlie interferes modes A$_{\text{2}}$ and B$_{\text{2}}$ coming from Alice and Bob respectively at a 50:50 beamsplitter and performs single-photon detection with PNRDs. A successful outcome occurs when Charlie's left (10 event) or right detector (01 event) registers a single click only. Alice and Bob ignore the instances of 02, 20 and 11 photon events. (b) Prepare-and-measure scheme where Alice and Bob send single-mode states to Charlie where they encode the key information in the $Z$ basis. They send the states $\ket{0}$ and $\ket{1}$ with probabilities $a_0, b_0$ and $a_1, b_1$, respectively. They randomly switch to $X$ and $Y$ bases to send check states to estimate Eve's information (refer to Sec.~\ref{sec:checkstatessection} for the details of the check states). Charlie's measurement is the same as the one in the entanglement-based scheme shown in panel (a).}
\end{figure*}

In Fig.~\ref{fig:repeater_plots}(b), we show the RCI of the optimised states with \text{$n_\text{max}=7$} and \text{$n_\text{max}=1$} when Alice and Bob perform point-to-point and single-repeater communications. We compare these results with the PLOB bound as it coincides with the RCI of a maximally entangled TMSV state in the pure-loss channel. Note that when Alice and Bob communicate directly using the optimised states with \text{$n_\text{max}=7$}, they cannot saturate the PLOB bound due to sending states with a limited number of photons. However, they can reach the PLOB bound if they send infinitely squeezed TMSV states with an infinite number of photons~\cite{pirandola2020advances}. In Fig.~\ref{fig:repeater_plots}(b), when Alice and Bob perform point-to-point communication, they can distill more entanglement at short distances. However, with the use of a repeater, they are able to distill more entanglement beyond $47$ km and surpass the RCI of an infinitely squeezed TMSV state at $108$ km. Note that they also surpass the PLOB bound at this distance when we calculate their secret key rate  as shown in Fig.~\ref{fig:repeater_plots}(a) and the key rates coincide with the reverse coherent information of Alice and Bob's conditional joint state on Charlie's measurement outcome. This indicates that after Charlie's measurement, Alice and Bob's PNRD measurement is optimal as Alice and Bob achieve the same key rates as the distillable entanglement of their joint state.

\subsection{\label{sec:realimplementation}Realistic Implementation of the MDI Protocol with Single-Photon States}
The experimental realisation of the higher dimensional optimised states and Charlie's measurement is quite challenging with state of the art technology. However, we present an experimentally feasible implementation of our protocol, shown in Fig.~\ref{fig:single_photon_experiment}, by using single-photon states which can be performed with existing technology. Fig.~\ref{fig:repeater_plots}(a) demonstrates that beyond $10$ km, the single-photon states achieve the same key rates as the higher dimensional states and the high-dimensional states reduce down to the single-photon level as demonstrated in Fig.~\ref{fig:opt_coefficients_figure}(a) and Fig.~\ref{fig:opt_coefficients_figure}(b).
\begin{figure}[t!]
\hspace*{-0.3cm}
\includegraphics[scale=0.62]{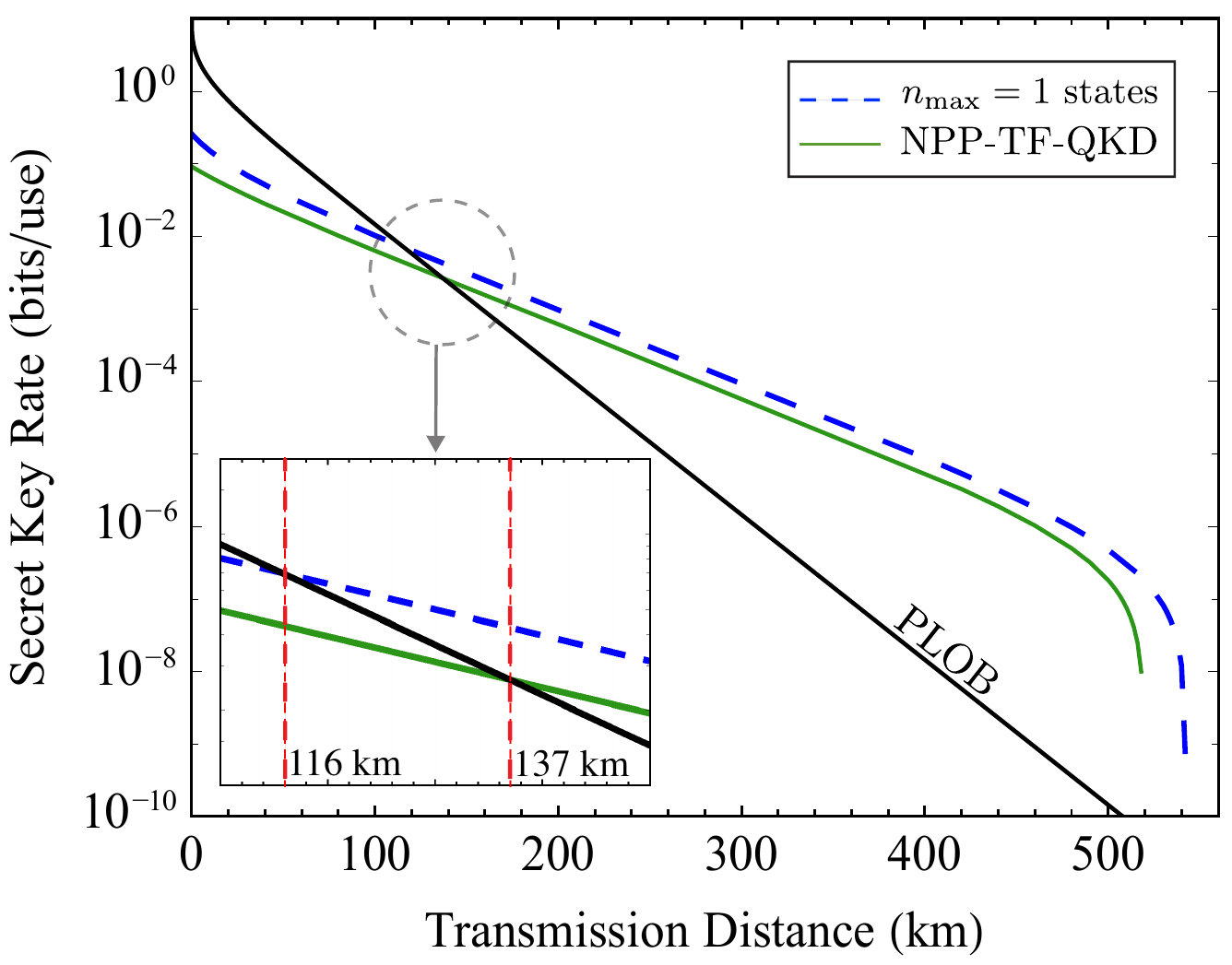}
\caption{\label{fig:single_photon_plots} The simulation results of our repeater protocol using single-photon states with experimental parameters. The blue dashed line shows our protocol using the single-photon states in the form of \text{${\sqrt{a_0}}\ket{00}+{\sqrt{a_1}}\ket{11}$} and \text{${\sqrt{b_0}}\ket{00}+{\sqrt{b_1}}\ket{11}$} with optimised coefficients. The solid green line and black lines are the NPP-TF-QKD protocol and the PLOB bound, respectively. We assume a dark count rate of \text{$5\times10^{-8}$} with a detector efficiency of $0.85$ for both protocols.}
\end{figure}

When Alice and Bob send single-photon states, Charlie can measure from $0$ to $2$ photons. However, as mentioned previously in Sec.~\ref{sec:calkeyrate}, when Charlie measures $2$ photons, the conditional secret key rate is zero as such the contribution to the key rate comes from only the single-photon detection events. This eliminates Charlie having to distinguish between $c=2$ photon outcomes, i.e., \text{$\ket{\phi_{2}^0}$}, \text{$\ket{\phi_{2}^1}$} and \text{$\ket{\phi_{2}^2}$} and requires him to only distinguish between the single-photon outcomes. Therefore, we can simplify our protocol to Fig.~\ref{fig:single_photon_experiment}, where Charlie interferes the single photons coming from Alice and Bob at a 50:50 beamsplitter and uses two photon-number resolving detectors up to the two-photon level. After Charlie's measurement, Alice and Bob can estimate their joint state to bound Eve's information using the statistics of their matched and unmatched data of $X$, $Y$ and $Z$ bases as mentioned in Sec.~\ref{sec:calkeyrate}.

Additionally, we consider the detrimental effects of the detector inefficiency and dark counts to the key rates. The single-photon states are optimised for a detector with an efficiency of $85\%$ and a dark count rate of $5\times10^{-8}$ where the coefficients of the zero and single photons are shown in Table~\ref{tab:singlephotontable} and in Fig.~\ref{fig:opt_coefficients_figure}(b). In a lossless channel, the probability of sending a single-photon initially is half. However, as the channel becomes more lossy, it is likely that the single-photon will be lost during transmission. When Charlie receives no photons, this corresponds to a large bit-error rate reducing the key rates. This is compensated by reducing the probability of sending single-photons to decrease the bit-error rates and increase the key rates~\cite{yin2019measurement}.

With realistic dark count rates and detector efficiencies, our protocol surpasses the PLOB bound at $116$ km while the NPP-TF-QKD surpasses at $137$ km as shown in Fig.~\ref{fig:single_photon_plots}. The NPP-TF-QKD protocol drops to zero beyond $518$ km whereas our protocol drops to zero beyond $542$ km showing a $24$ km advancement in the transmission distance. These improvements are a result of several factors. Even though both protocols use optimised states, our protocol has more freedom over optimising the coefficients of the single-photon state while the TF-QKD protocols need to ensure that the intensities of the coherent states are still weak enough while optimising the key rates. This is also one of the key differences between our protocol and the Sending-or-Not-Sending TF-QKD (SNS-TF-QKD) protocol~\cite{wang2018twin}, where Alice and Bob send weak coherent states and no states with a probability of \text{$\epsilon$} and \text{$1-\epsilon$}, respectively. However, the probability of the single-photon detection is still determined by the intensity of the weak coherent states in the SNS-TF-QKD protocol whereas in this protocol, Alice and Bob send single-photon states with a probability of \text{$\epsilon$} which determines the probability of detection at Charlie's detectors. Our protocol also has the ability to distinguish two-photon events occurring at a single detector at Charlie. For example, if Charlie receives no photons on one detector and two photons on the other, these events can be disregarded and do not contribute to bit-error rates. However, in TF-QKD protocols with single-photon detectors, this event would register as one click, causing an increase in the bit-error rate. Therefore, the use of PNRDs in Charlie's station improves the bit-error rates. Furthermore, our protocol can estimate Eve's information more accurately due to the use of the probabilities of the matched and unmatched bases. These are the main factors that distinguish our protocol from the existing MDI and TF-QKD protocols.


\section{\label{sec:level5}Discussion}
In this paper, we introduced a new MDI protocol using higher dimensional states that surpasses the repeaterless bound without the need of quantum memories as it scales like the single-repeater bound. However, for large distances, the states required in this protocol reduce down to the single-photon level due to the losses in the channel. Based on this, we proposed an experimentally feasible implementation of this protocol just using single-photons and photon-number resolving detectors which performs better than the existing protocols such as NPP-TF-QKD protocol~\cite{cui2019twin,lu2019improving}. 

Furthermore, we investigated whether the single-repeater bound can be saturated with a simple protocol by using only single copies of the states sent by Alice and Bob and without collective measurements performed by Charlie. Our results show that unlike the repeaterless bound, this is probably not possible with single copies of the states and likely to require many copies of the states sent by Alice and Bob and collective measurements as previously shown by Garc\'{i}a-Patr\'{o}n et al.~\cite{garcia2009reverse} and a new protocol proposed by Winnel et al.~\cite{winnel2022achieving}.

The results presented in this work refer to the asymptotic key rates, and the security of this protocol with finite-size effects needs to be considered in the future. In this protocol, there are no misalignment errors in the $Z$ basis due to sending single-photons. However, the misalignment errors are likely to impact the statistics of the check states in the $X$ and $Y$ bases which can be investigated in future work. The feasibility of extending this protocol to a network of multiple users can also be studied.

\section{Methods}
\subsection{\label{sec:appDVtomogprahy} Estimating Eve's Information Using Quantum Tomography with Single-Photon States}
In this section, we show how Alice and Bob can estimate their joint state conditioned on Charlie's measurement outcome to bound Eve's information. 

Alice and Bob measure their joint state in the $X$, $Y$ and $Z$ bases in the entanglement-based scheme as introduced in Sec.~\ref{sec:checkstatessection} to construct the statistics of their matched and unmatched results. From the probabilities measured in these bases, Alice and Bob can estimate their joint state. Writing their joint state as
\begin{equation}
   \hat{\rho}_{\text{AB}|c=1}=\frac{1}{4}\big(\mathbb{I}_{4}+\vec{s}_a\!\cdot\vec{\sigma}_a+\vec{s}_b\cdot\vec{\sigma}_b+\sum_{j,k}{r_{jk}}(\sigma_{j}{\otimes}\sigma_{k})\big),
\label{eq:fanorep}
\end{equation}
where \text{$\vec{s}_a\!\cdot\vec{\sigma}_a$} and \text{$\vec{s}_b\!\cdot\vec{\sigma}_b$} describe Alice and Bob's reduced states calculated from their local measurements while \text{$r_{jk}(\sigma_{j}{\otimes}\sigma_{k})$} gives the correlations between Alice and Bob determined from their measurements performed in the bases \text{$j=\{X,Y,Z\}$} and \text{$k=\{X,Y,Z\}$}
where \text{$r_{jk}$} is the correlation coefficient and \text{$\sigma_{j}$} and \text{$\sigma_{k}$} are the standard Pauli matrices \text{$\sigma_X$}, \text{$\sigma_Y$} and \text{$\sigma_Z$}. The terms \text{$\vec{s}_a\cdot\vec{\sigma}_a$} and \text{$\vec{s}_b\cdot\vec{\sigma}_b$} can be expressed as
\begin{subequations}
\label{eq:reducedstates}
\begin{equation}
\label{eq:alicesqubit}
\vec{s}_a\!\cdot\vec{\sigma}_a=a_X(\sigma_X{\otimes}\mathbb{I}_{2})+a_Y(\sigma_Y{\otimes}\mathbb{I}_{2})+a_Z(\sigma_Z{\otimes}\mathbb{I}_{2}),
\end{equation}
\begin{equation}
\label{eq:bobsqubit}
\vec{s}_b\!\cdot\vec{\sigma}_b=b_X(\mathbb{I}_{2}{\otimes}\sigma_X)+b_Y(\mathbb{I}_{2}{\otimes}\sigma_Y)+b_Z(\mathbb{I}_{2}{\otimes}\sigma_Z).
\end{equation}
\end{subequations}
where $\{a_X,a_Y,a_Z\}$ and $\{b_X,b_Y,b_Z\}$ represent the coefficients given in Eq.~\eqref{eq:abcoeffs} when Alice and Bob measure in the bases $j=\{X,Y,Z\}$.

When Alice and Bob measure their own qubits in any basis \text{$j=\{X,Y,Z\}$}, their measurement outcomes can be expressed as
\begin{equation}
    \Pi_{\pm j}=\ketbra{\pm j},
\end{equation}
which can be calculated using the eigenvectors of the $X$, $Y$ and $Z$ bases as defined previously in Eq.~\eqref{eq:allbases}. The probability of their measurement then can be calculated from
\begin{subequations}
\begin{equation}
    P_{a}(\pm j)=\mathrm{Tr}\big[\big(\Pi_{\pm j}{\otimes} \mathbb{I}_{2}\big)\rho_{\text{AB}|c=1}\big(\Pi_{\pm j}{\otimes} \mathbb{I}_{2}\big)^\dagger\big],
\label{eq:palice}
\end{equation}
\begin{equation}
    P_{b}(\pm j)=\mathrm{Tr}\big[\big(\mathbb{I}_{2}{\otimes}\Pi_{\pm j}\big)\rho_{\text{AB}|c=1}\big(\mathbb{I}_{2}{\otimes}\Pi_{\pm j}\big)^\dagger\big],
\label{eq:pbob}
\end{equation}
\end{subequations}
where \text{$\rho_{AB|c=1}$} is determined from Eq.~\eqref{eq:rhoABnorm}.

The coefficients in Eqs.~\eqref{eq:alicesqubit} and~\eqref{eq:bobsqubit} can computed from Eqs.~\eqref{eq:palice} and~\eqref{eq:pbob} where
\begin{subequations}
\label{eq:abcoeffs}
\begin{equation}
a_j=P_{a}(+j)-P_{a}(-j),
\label{eq:aj}
\end{equation}
\begin{equation}
b_j=P_{b}(+j)-P_{b}(-j).
\label{eq:bj}
\end{equation}
\end{subequations}

In order to determine the correlation coefficients \text{$r_{jk}$}, Alice and Bob construct a joint probability table of their measurements in all the bases where these probabilities are calculated from
\begin{equation}
P(a\!=\!\pm j,b\!=\!\pm k){=}\mathrm{Tr}\big[\!\big(\Pi_{\pm j}{\otimes}\Pi_{\pm k}\big)\rho_{\text{AB}|c=1}\big(\Pi_{\pm j}{\otimes}\Pi_{\pm k}\big)\!^\dagger\big].
\label{eq:probrjk}
\end{equation}
Using Eq.~\eqref{eq:probrjk}, the correlation coefficients become
\begin{multline}
    r_{jk}=P(a\!=\!+j,b\!=\!+k)+P(a\!=\!-j,b\!=\!-k)\\-P(a\!=\!+j,b\!=\!-k)-P(a\!=\!-j,b\!=\!+k).
    \label{eq:correlations}
\end{multline}
After Alice and Bob reconstruct their estimated joint matrix \text{$\hat{\rho}_{\text{AB}|c=1}$}, they can estimate Eve's information using the Holevo bound as given in Eq.~\eqref{eq:eveinfo}.

\subsection{\label{sec:appkeyclassical}Classical Protocol Used to Optimise the Coefficients of the High Dimensional States}
This section describes how the states that Alice and Bob prepare are chosen. The coefficients of these states are determined based on the following classical protocol. We assume Eve taps off the signal sent by Alice and Bob, and measures the number of photons denoted as \text{$n_{e_a}$} and \text{$n_{e_b}$}. Then we maximise the average difference in mutual information
\begin{equation}
\max\limits_{\{P(n_a),P(n_b)\}}\bigg[\sum_{c=0}^{2n_\text{max}}P_{c}(n_c)\big(I_{AB|c}-I_{AE|c}\big)\bigg],
\label{eq:diffmutualinfo}
\end{equation}
where \text{$I_{AB|c}$} and \text{$I_{AE|c}$} are Alice and Bob's mutual information and mutual information between Alice and Eve conditioned on Charlie's measurement outcome respectively. $P(n_c)$ represents the probability of Charlie measuring $n_c$ photons in total. Note that $P(n_a)$ and $P(n_b)$ are related to the optimised coefficients from Eq.~\eqref{eq:statealice} as they are the probability of sending \text{$n$} photons for the corresponding Fock-number state \text{$\ket{n}$}, also expressed as $a_n$ and $b_n$ throughout this paper.

In the classical protocol, Charlie measures the number of photons coming from Alice and Bob individually with two separate PNRDs. In Fock basis, both classical and quantum simulations yield the same probabilities for Charlie's measurement outcome. The probability of Charlie measuring \text{$n_{c_a}$} or \text{$n_{c_b}$} photons on Alice's and Bob's mode individually can be computed as
\begin{equation}
P_{c_a}(n_{c_{a}})=\sum_{n_{a}\!=0}^{n_{\text{max}}}\binom{n_{a}}{n_{c_{a}}}\tau_{\text{A}}^{n_{c_a}}(1-\tau_{\text{A}})^{n_{a}-n_{c_{a}}}P({n_{a})},
\label{eq:probsending}
\end{equation}
where \text{$n_\text{max}$} refers to the maximum number of photons Alice and Bob are sending individually. \text{$\tau_{\text{A}}$} is the probability of a photon arriving at Charlie from Alice or Bob as a function of the fibre distance with \text{$\tau_{\text{A}}=10^{-0.02d}$}. \text{$(1-\tau_{\text{A}})^{n_{a}-n_{c_{a}}}$} is the probability of losing \text{$n_{a}-n_{c_{a}}$} photons to Eve. The probability of the collective photon number measurement performed by Charlie for a given number of photons $n_c$ can be calculated using Eq.~\eqref{eq:probsending} as shown below
\begin{equation}
P_{c}(n_{c})=\sum_{n_{c_{a}}\!=0}^{n_{c}}P_{c_{a}}(n_{c_{a}})P_{c_{b}}(n_{c}-n_{c_{a}}),
\label{eq:probcharlie}
\end{equation}
where $n_{c}-n_{c_{a}}$ gives the number of photons measured on Bob's mode.

Alice and Bob's mutual information conditioned on Charlie's measurement outcome is obtained from the probability table between Alice and Bob which is as follows
\begin{equation}
P(n_{a},n_{b}|n_{c}){=}{\binom{\!n_{a}+n_{b}\!}{\!n_c\!}}\!\frac{\tau^{n_{c}}(1-\tau)^{n_{a}+n_{b}-n_c}P(n_a)P(n_b)}{P_{c}(n_{c})},
\label{eq:probtableAB}
\end{equation}
where \text{$n_a+n_b$} is equal to the total number of photons in the system and \text{$\tau$} in the equation above corresponds to the transmission probability in one channel only. We evaluate Alice and Bob's mutual information conditioned on Charlie's measurement outcome from the same approach shown in Sec.~\ref{sec:calkeyrate} using \text{$I_{AB|c}=H(A|c)+H(B|c)-H(AB|c)$} and Eqs.~\eqref{eq:subeqsalice}, \eqref{eq:subeqsbob} and \eqref{eq:subeeqsabtogether}.

We quantify Eve's information conditioned on each photon measurement in a similar fashion as Alice and Bob's mutual information using \text{$I_{E|c}=H(A|c)+H(E|c)-H(AE|c)$}. Since Eve has access to both channels between Alice and Charlie and Charlie and Bob, we need to consider events where each party loses photons to Eve. We compute the probability table between Alice, Bob and the two modes of Eve conditioned on Charlie's outcome as follows
\begin{multline}
    P(n_a,n_b,n_{e_a},n_{e_b}|n_c)=
    \frac{1}{P_c(n_{c})}\binom{n_{a}}{n_{e_{a}}}\binom{n_{b}}{n_{e_{b}}}\\\tau^{n_{a}+n_{b}-(n_{e_{a}}+n_{e_{b}})}(1-\tau)^{n_{e_{a}}+n_{e_{b}}}P(n_a)P(n_b),
    \label{eq:probtableABEve}
\end{multline}
provided \text{$n_{a}+n_{b}-(n_{e_{a}}+n_{e_{b}})=n_{c}$}, where \text{$n_{e_a}$} and \text{$n_{e_b}$} are the photons lost to Eve by Alice and Bob respectively and \text{$n_{a}+n_{b}-(n_{e_{a}}+n_{e_{b}})$} corresponds to the total number of photons measured by Charlie. Therefore, using the probability table between Alice, Bob and Eve, we can calculate the entropies below to compute Eve's information
\begin{subequations}
\begin{multline}
H(E_\text{A}E_\text{B}|c)=-\sum_{n_{e_{a}}\!=0}^{n_{\text{max}}}\sum_{n_{e_{b}}\!=0}^{n_{\text{max}}}\\P(n_{e_a},n_{e_b}|c)\log_{2}P(n_{e_a},n_{e_b}|c),
\label{eq:eqse1e2}
\end{multline}
\begin{multline}
H(AE_\text{A}E_\text{B}|c)=-\sum_{n_{a}\!=0}^{n_{\text{max}}}\sum_{n_{e_a}\!=0}^{n_{\text{max}}}\sum_{n_{e_b}\!=0}^{n_{\text{max}}}P(n_{a},n_{e_{a}},n_{e_{b}}|c)\\\log_{2}P(n_{a},n_{e_{a}},n_{e_{b}}|c).
\label{eq:eqsae}
\end{multline}
\end{subequations}

\subsection{\label{sec:darknoisemodel}Modelling Dark Noise in the Entanglement Swapping Measurement}
\begin{figure}[htp]
        \center{\includegraphics[width=0.33\textwidth]
        {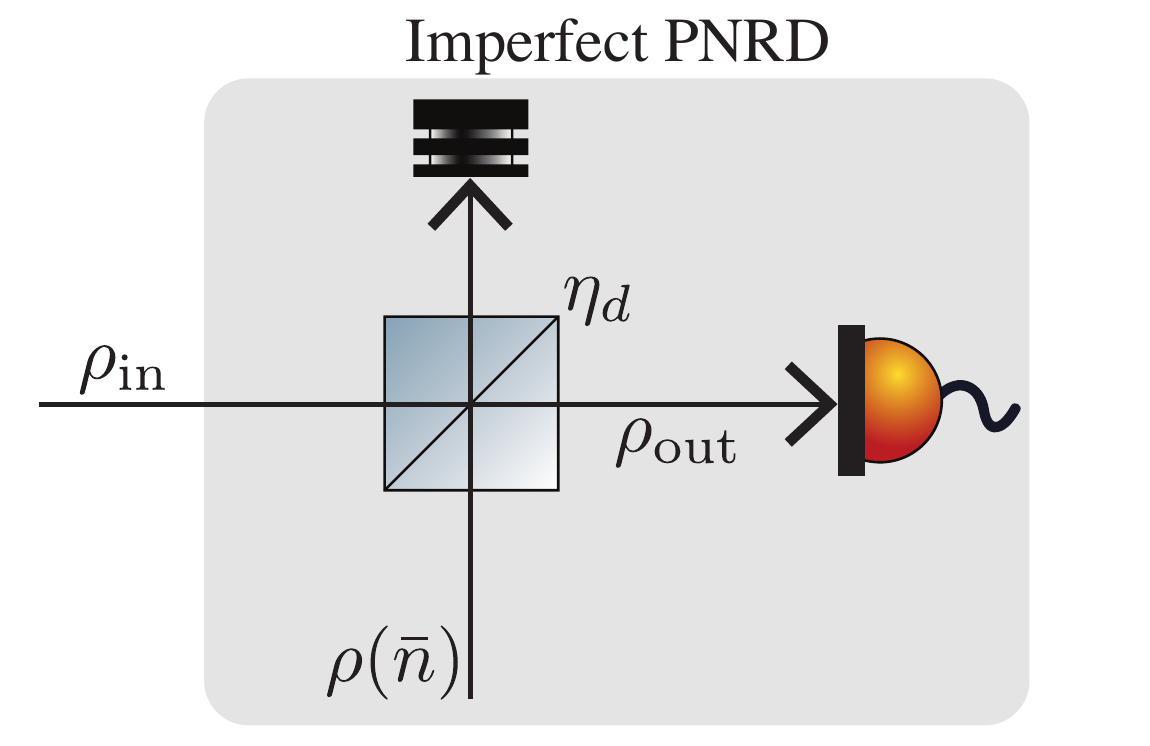}}
        \caption{\label{PNRD} Diagram of the method used to simulate the effect of dark counts in the PNRDs. $\rho_{\text{in}}$ is the density matrix of the input state. The grey box represents a realistic photon number resolving detector with efficiency $\eta_d$ and dark noise $(1-\eta_d)\bar{n}$.}
\end{figure}
This section describes how to model the dark noise and detector efficiency at Charlie's photon detectors to achieve the results of Fig.~\ref{fig:single_photon_plots}. The effects of dark noise is modelled by interacting the incoming state with a thermal state at a beamsplitter as in Fig.~\ref{PNRD}.

The efficiency of the single photon detection in this framework is the transmissivity of the beamsplitter~($\tau$), i.e., $\eta_d = \tau$. The density matrix of the state to be detected can be written down as
\begin{equation}
\rho_{\text{out}} = B(\eta_d)(\rho_{\text{in}}{\otimes}\rho(\bar{n}))B(\eta_d)^{\dagger},
\end{equation}
where $\rho(\bar{n})$ is the density matrix of the thermal state. The beamsplitter transformation is shown in Eq.~\eqref{eq:beamsplitter}. The density matrix of the thermal state is given by
\begin{equation}
\label{eq:thermalstdark}
\rho(\bar{n}) = \sum_{n=0}^{\infty} \frac{\bar{n}^n}{(1+\bar{n})^{n+1}}\ketbra{n},
\end{equation}
where $\bar{n}=\text{Tr}[\rho(\bar{n})a^{\dagger}a]$ is the mean photon number of the thermal state. Consequently the dark count is given by $(1-\eta_d)\bar{n}$. For low dark counts, the summation in Eq.~\eqref{eq:thermalstdark} can be truncated accordingly.

\subsection{\label{sec:appcoeffs} Coefficients of the Optimised States}
In this section, we present some of the coefficients of the optimised states with \text{$n_\text{max}=7$} and \text{$n_\text{max}=1$} photons used in Fig.~\ref{fig:repeater_plots}(a) and (b) for each distance in Tables~\ref{tab:cvtable} and~\ref{tab:dvtable} correspondingly. These coefficients represent the probability of sending the corresponding Fock-number state. We give the values of the optimised squeezing parameters of the states given in Eq.~\eqref{eq:eprstate} with \text{$n_\text{max}=7$} photons for each distance used in Fig.~\ref{fig:repeater_plots}(a) in Table~\ref{tab:eprtable}. In Table~\ref{tab:singlephotontable}, we present the coefficients of the optimised single-photon states shown in Fig.~\ref{fig:single_photon_plots}.

\begin{table*}[htp!]
\caption{\label{tab:cvtable}
The coefficients of the optimised states with  \text{$n_\text{max}=7$} photons.}
\begin{ruledtabular}
\begin{tabular}{P{0.1\linewidth}|c|c|c|c|c|c|c|c}
\makecell{Distance \\ (km)} & \makecell{\text{$a_0$} \\ \text{$b_0$}} & \makecell{\text{$a_1$} \\ \text{$b_1$} } & \makecell{\text{$a_2$} \\ \text{$b_2$} } & \makecell{\text{$a_3$} \\ \text{$b_3$}} & \makecell{\text{$a_4$} \\ \text{$b_4$}} & \makecell{\text{$a_5$} \\ \text{$b_5$}} & \makecell{\text{$a_6$} \\ \text{$b_6$}} & \makecell{\text{$a_7$} \\ \text{$b_7$}}\\
\colrule
$0$ & $0.0823$ & $0.1162$ & $0.1432$ & $0.1582$ & $0.1582$ & $0.1432$ & $0.1162$ & $0.0823$\\
$0.5$ & $0.1073$ & $0.1359$ & $0.1557$ & $0.1603$ & $0.1496$ & $0.1267$ & $0.0968$ & $0.0676$\\
$1$ & $0.1226$ & $0.1472$ & $0.1616$ & $0.1600$ & $0.1438$ & $0.1174$ & $0.0868$ & $0.0605$\\
$2.5$ & $0.1550$ & $0.1705$ & $0.1710$ & $0.1567$ & $0.1307$ & $0.0994$ & $0.0691$ & $0.0477$\\
$5$ & $0.1967$ & $0.2012$ & $0.1786$ & $0.1483$ & $0.1129$ & $0.0787$ & $0.0508$ & $0.0329$\\
$10$ & $0.4468$ & $0.3137$ & $0.1410$ & $0.0601$ & $0.0245$ & $9.4433\times10^{-3}$ & $3.3895\times10^{-3}$ & $1.1308\times10^{-3}$\\
$15$ & $0.6654$ & $0.2955$ & $0.0366$ & $2.4273\times10^{-3}$ & $1.1213\times10^{-4}$ & $3.9036\times10^{-6}$ & $9.8873\times10^{-8}$ & $0$\\
$20$ & $0.7230$ & $0.2608$ & $0.0160$ & $2.8606\times10^{-4}$ & $2.1895\times10^{-6}$ & $3.2584\times10^{-9}$ & $0$ & $0$ \\
$25$ & $0.7548$ & $0.2366$ & $8.5496\times10^{-3}$ & $4.9545\times10^{-5}$ & $6.2159\times10^{-8}$ & $0$ & $0$ & $0$ \\
$30$ & $0.7760$ & $0.2189$ & $5.0283\times10^{-3}$ & $1.0464\times10^{-5}$ & $0$ & $0$ & $0$ & $0$ \\
$50$ & $0.8176$ & $0.1811$ & $1.3468\times10^{-3}$ & $1.2642\times10^{-7}$ & $0$ & $0$ & $0$ & $0$ \\
$100$ & $0.8477$ & $0.1520$ & $3.3582\times10^{-4}$ & $0$ & $0$ & $0$ & $0$ & $0$ \\
$200$ & $0.8571$ & $0.1427$ & $1.9467\times10^{-4}$ & $0$ & $0$ & $0$ & $0$ & $0$ \\
\end{tabular}
\end{ruledtabular}
\end{table*}

\begin{table}[htp!]
\caption{\label{tab:dvtable}
The coefficients of the optimised states with  \text{$n_\text{max}=1$} photon.} 
\begin{ruledtabular}
\begin{tabular}{P{2.5cm}|P{2.7cm}|P{2.7cm}}
\makecell{Distance \\ (km)} & \makecell{\text{$a_0$} \\ \text{$b_0$}} & \makecell{\text{$a_1$} \\ \text{$b_1$}}\\
\colrule
$0$ & $0.5$ & $0.5$ \\
$0.5$ & $0.5308$ & $0.4692$ \\
$1$ & $0.5505$ & $0.4495$ \\
$2.5$ & $0.5918$ & $0.4082$ \\
$5$ & $0.6371$ & $0.3629$ \\
$10$ & $0.6935$ & $0.3065$ \\
$15$ & $0.7292$ & $0.2708$ \\
$20$ & $0.7542$ & $0.2458$ \\
$30$ & $0.7869$ & $0.2131$ \\
$50$ & $0.8205$ & $0.1795$ \\
$100$ & $0.8483$ & $0.1517$ \\
$200$ & $0.8575$ & $0.1425$ \\
\end{tabular}
\end{ruledtabular}
\end{table}

\begin{table}[htp!]
\caption{\label{tab:eprtable}
The optimised squeezing parameters ($\gamma$) of the states shown in Eq.~\eqref{eq:eprstate} with \text{$n_\text{max}=7$} photons.
}
\begin{ruledtabular}
\begin{tabular}{P{4cm}|P{4.2cm}}
Distance (km) & Squeezing Parameter (\text{$\gamma$}) \\
\colrule
$0$ & $0.84$ \\
$0.5$ & $0.83$ \\
$1$ & $0.83$ \\
$2.5$ & $0.82$ \\
$5$ & $0.81$ \\
$10$ & $0.71$ \\
$15$ & $0.52$ \\
$20$ & $0.44$ \\
$25$ & $0.40$ \\
$30$ & $0.37$ \\
$40$ & $0.33$ \\
$50$ & $0.30$ \\
$100$ & $0.26$ \\
$200$ & $0.25$ \\
\end{tabular}
\end{ruledtabular}
\end{table}

\begin{table}[h!]
\caption{\label{tab:singlephotontable}
The coefficients of the single-photon states when the detector dark count rate is \text{$5\times10^{-8}$} and with a detector efficiency of $0.85$.
}
\begin{ruledtabular}
\begin{tabular}{P{2.5cm}|P{2.7cm}|P{2.7cm}}
\makecell{Distance \\ (km)} & \makecell{\text{$a_0$} \\ \text{$b_0$}} & \makecell{\text{$a_1$} \\ \text{$b_1$}}\\
\colrule
$0.5$ & $0.6697$ & $0.3303$ \\
$1$ & $0.6751$ & $0.3249$ \\
$2.5$ & $0.6896$ & $0.3104$ \\
$5$ & $0.7100$ & $0.2900$ \\
$10$ & $0.7405$ & $0.2595$ \\
$15$ & $0.7624$ & $0.2376$ \\
$20$ & $0.7790$ & $0.2210$ \\
$30$ & $0.8020$ & $0.1980$ \\
$50$ & $0.8275$ & $0.1725$ \\
$100$ & $0.8499$ & $0.1501$ \\
$200$ & $0.8576$ & $0.1424$ \\
$400$ & $0.8588$ & $0.1412$ \\
$420$ & $0.8591$ & $0.1409$ \\
$440$ & $0.8598$ & $0.1402$ \\
$460$ & $0.8609$ & $0.1391$ \\
$480$ & $0.8630$ & $0.1370$ \\
$490$ & $0.8647$ & $0.1353$ \\
$500$ & $0.8669$ & $0.1331$ \\
$516$ & $0.8721$ & $0.1279$ \\
$518$ & $0.8730$ & $0.1270$ \\
$520$ & $0.8739$ & $0.1261$ \\
$522$ & $0.8749$ & $0.1251$ \\
$524$ & $0.8760$ & $0.1240$ \\
$530$ & $0.8796$ & $0.1204$ \\
$532$ & $0.8810$ & $0.1190$ \\
$534$ & $0.8825$ & $0.1175$ \\
$536$ & $0.8841$ & $0.1159$ \\
$538$ & $0.8859$ & $0.1141$ \\
$540$ & $0.8878$ & $0.1122$ \\
$542$ & $0.8898$ & $0.1102$ \\
\end{tabular}
\end{ruledtabular}
\end{table}

\section*{Data availability}
The data that supports the findings of this study are available from the corresponding author upon reasonable request.

\section*{Code availability}
The codes that support the findings of this study are available from the corresponding author upon reasonable request.

\FloatBarrier

\bibliography{apssamp}

\bibliographystyle{naturemag}

\section*{Acknowledgments}
We thank Matthew S. Winnel for his valuable discussion during this project. This research was funded by the Australian Research Council Centre of Excellence for Quantum Computation and Communication Technology (Grant No. CE110001027). Y.-S.K acknowledges support from the KIST institutional program (2E31021).
\vspace{0.6cm}

\section*{Author contributions}
O.E. conceived the project. O.E. and S.A. developed the theory. O.E. performed the numerical analysis. O.E. wrote the manuscript. All authors contributed towards~the~theory,~discussions~of~the~results~and~the manuscript. S.A. supervised the project.

\vspace{-0.1cm}
\section*{Competing Interests}
The authors declare no competing financial or non-financial interests.

\end{document}